\documentclass{aa} 
\usepackage{graphicx} 
\usepackage{times}
\usepackage{natbib} 
\bibpunct{(}{)}{;}{a}{}{,} 

\newcommand{\msuns}{$\mathrm{{M}_\odot}$ }
\newcommand{\mjup}{$\mathrm{{M}_J}$}
\newcommand{\mjups}{$\mathrm{{M}_J}$ }

\begin{document}
  \title{Planets opening dust gaps in gas disks} 
  \author{Sijme-Jan Paardekooper \and Garrelt Mellema} 
  \author{Sijme-Jan Paardekooper \inst{1} \and Garrelt Mellema \inst{2,1}} 
  \offprints{S. J. Paardekooper\\\email{paardeko@strw.leidenuniv.nl}} 
  \institute{Leiden Observatory, Postbus 9513, NL-2300 RA Leiden, 
             The Netherlands \and
             ASTRON, Postbus 2, NL-7990 AA Dwingeloo, The Netherlands \\
    \email{paardeko@strw.leidenuniv.nl; gmellema@astron.nl}} 
  \date{Received ; Accepted}
  
  \abstract{We investigate the interaction of gas and dust in a protoplanetary
    disk in the presence of a massive planet using a new two-fluid 
    hydrodynamics code. In view of future observations of planet-forming 
    disks we focus on the condition for gap formation in the dust fluid. 
    While only planets more massive than 1 Jupiter mass (\mjup) open up a 
    gap in the gas disk, we find that a planet of 0.1 \mjups already creates
    a gap in the dust disk. This makes it easier to find lower-mass planets 
    orbiting in their protoplanetary disk if there is a significant population 
    of mm-sized particles.
     
     } 
  
  \maketitle
  
\section{Introduction}
Resolved images of protoplanetary disks can reveal density structures in 
gas and dust hinting at the presence of embedded planets. It is therefore 
important to investigate the response of a dusty circumstellar disk on 
planets, as opposed to for example radiation pressure effects in optically
thin disks \citep{2001ApJ...557..990T}.

The most dramatic disk structure a planet can create is an annular gap
\citep{1984ApJ...285..818P}, and these planet-induced gaps can be used 
as tracers for recent giant planet formation. 
Recently, \cite{2002ApJ...566L..97W} showed that the Atacama Large
Millimeter Array (ALMA) will be able to detect a gap at 5.2 AU caused
by the presence of a Jupiter mass planet in a disk 140 pc away in the
Taurus star-forming region. However, they assumed a constant dust-to-gas
mass ratio of 1:100, whereas in the presence of large pressure gradients
the gas and the dust evolution will decouple to some extent 
\citep{2002ApJ...581.1344T}. Due to radial pressure support
the gas orbits at sub-Keplerian velocity, and when the dust particles are 
slowed down by drag forces they move inward.

In this letter we present the first results of two-dimensional numerical
simulations where we treat the gas and the dust as separate but coupled 
fluids. These two-fluid calculations take into account the full interaction 
of gas and dust, and can be used to simulate observations in a better way.

In Sect. \ref{secequations} we briefly discuss the physics of gas-dust
interaction. We describe the numerical method in Sect. \ref{secmethod},
and the initial conditions in Sect. \ref{secinitial}. We present the 
results in Sect. \ref{secres} and we conclude in Sect. \ref{secdisc}.   

\section{Basic equations}
\label{secequations}
Protoplanetary disks are fairly thin, i.e. the vertical thickness $H$
is small compared with the distance $r$ from the centre of the disk.
Typically we use $H/r = h = 0.05$. Therefore it is convenient to average
the equations of motion vertically, and work with vertically averaged
quantities only. The governing equations as well as the gas disk model are
the ones described in detail in \cite{1999MNRAS.303..696K}.  

\subsection{Dust}
We treat the dust as a pressureless fluid, and its evolution is governed
by conservation of mass and conservation of linear and angular momentum.
The major difference between the equations for the gas and the dust is the
absence of pressure in the latter. In this sense the dust fluid behaves
like a gas that is moving always with supersonic velocity. This implies that
near shock waves, where the gas goes from sonic to supersonic flow and large
density and velocity gradients are present, dust and gas behave in a very 
different way. 

The interaction between the gas and dust fluids occurs only through
drag forces. The nature of the drag force depends on the size of the
particles with respect to the mean free path of the gas molecules. We
consider only spherical particles with radius $s$, and for the particle
sizes we are interested in ($s \approx 1$ mm) we can safely use Epsteins
drag law \citep{2001ApJ...557..990T}. In the limit of small relative 
velocities of gas and dust compared to the local sound speed, the drag 
forces are written as:
\begin{equation}
\label{eqdrag}
\vec{f}_{\mathrm{d}}=
-\Sigma \frac{\Omega_\mathrm{K}}{T_\mathrm{s}} \left(\vec{v} - 
\vec{v}_{\mathrm{g}}\right)
\end{equation}
where $\Omega_\mathrm{K}$ is the Keplerian angular velocity, and $T_\mathrm{s}$
is the non-dimensional stopping time \citep{2002ApJ...581.1344T}: 
\begin{equation}
T_\mathrm{s}=\sqrt{\frac{\pi}{8}} \frac{\rho_\mathrm{p} s 
v_\mathrm{K}}{\rho_\mathrm{g} r c_\mathrm{s}} 
\end{equation}
Here, $\rho_\mathrm{p}$ is the particle internal density, $s$ is the radius
of the particle and $c_\mathrm{s}$ is the sound speed. We have used 
$\rho_\mathrm{p}=1.25$ $\mathrm{g~cm^{-3}}$, and the gas density 
$\rho_\mathrm{g}$ is found by using $\rho_\mathrm{g}=\Sigma_\mathrm{g}/2 H$.
During simulations we have always checked whether Eq. \ref{eqdrag} applied.

\begin{figure*}
\includegraphics[bb=0 5 510 240,width=17cm]{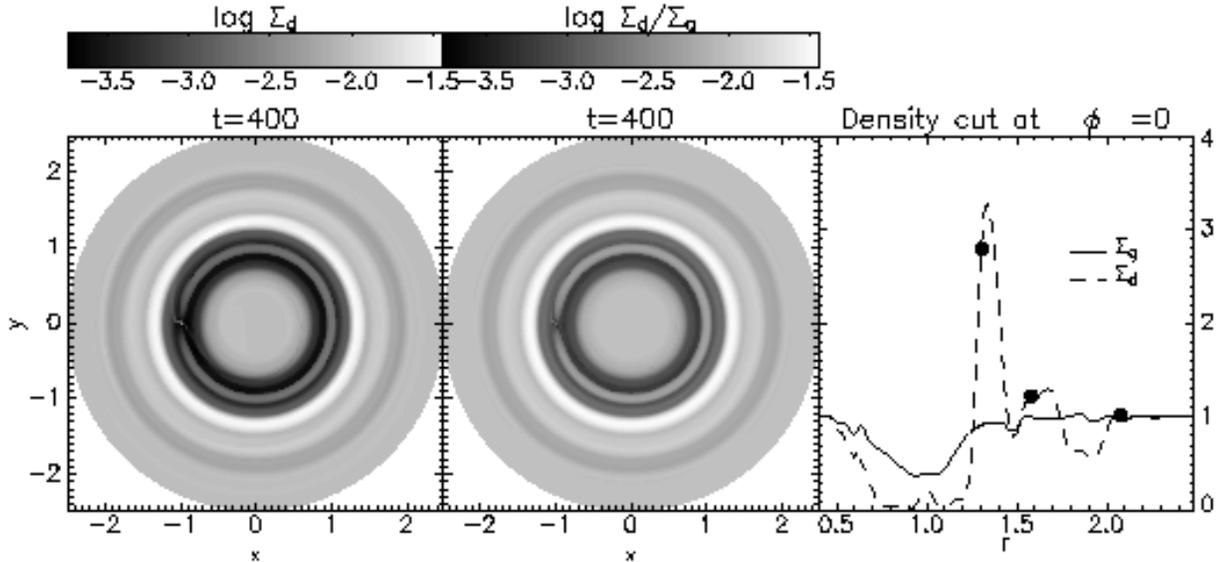}
\caption{Dust flow near a $0.1$ \mjups planet. Left panel: Grey-scale plot of 
the logarithm of the dust surface 
density after 400 planetary orbits. Middle panel: Logarithm of the
dust-to-gas ratio. Right panel: radial cut at $\phi=0$ (opposite to the
planet). Solid line: gas surface density, dashed line: dust surface density
$\times 100$. The filled circles indicate the 3:2, the 2:1 and the 3:1 mean 
motion resonances.}
\label{fig1}
\end{figure*}
\section{Numerical method}
\label{secmethod}
For the evolution of the gas component, we use the method described by
\cite{paardekooper03}. It is a second-order Eulerian hydrodynamics
code which uses an approximate Riemann solver \citep{1981...............}.
It derives from a general relativistic method \citep{1995A&AS..110..587E},
from which it inherits the capability to deal with non-Cartesian, non-inertial
coordinate systems. We work in a cylindrical coordinate frame $(r,\phi)$, 
corotating with the embedded planet that has a Keplerian angular velocity. 
A module for Adaptive Mesh Refinement (AMR) can be used
to obtain very high resolution near the planet. We use basically the same 
method to follow the evolution of the dust, but without the pressure terms.
This is a very different approach than for example in
\cite{1997Icar..128..213K}, who solve the equations of motion of particles 
instead of the fluid equations. All the source terms except the drag forces 
are integrated using stationary extrapolation \citep{paardekooper03}. The drag 
forces are incorporated separately using an ordinary differential equation:
\begin{equation}
\frac{d}{dt}(\Sigma \vec{v}) = \vec{f}_{\mathrm{d}} 
\end{equation}
Separating the drag forces from the advection in this way puts constraints
on the time step. Basically the assumption is that no drag force acts on the 
dust during the advection, and this is clearly not correct when the drag 
force source terms are very stiff. The time step is set by the hydrodynamic 
Courant-Friedrichs-Lewy (CFL) condition, and therefore one has to make sure 
that the typical time scale for the drag forces (the stopping time) is not
much smaller than this time step. Switching the order of integration
every time step yields a second order accurate update for the state 
\citep{1968...............}. If the stopping time is very small (well-coupled particles) the dust can not 
be assumed to move free of drag during a dust update. This puts 
a lower limit on the particle size that we can put into the simulations. For 
a typical disk, this limit is about 0.1 mm. The time step can however be 
reduced to include smaller particles.

\section{Initial and boundary conditions}
\label{secinitial}
The standard numerical resolution is $(n_r, n_\phi) = $ $(128,384)$.
This way, the cells close to the planet have equal size in the radial
and in the azimuthal direction. We do not resolve the Roche lobe at this
resolution, but since we are interested only in the large scale evolution 
of the disk, this is not important. We will discuss the flow within the 
Roche lobe and accretion processes in future papers. Using a low resolution 
allows us to run longer simulations, which is important because the 
drift velocities can be very low.
We take a constant initial surface density $\Sigma_0=34$ g $\mathrm{cm^{-2}}$.
This surface density corresponds to the value at $5.2$ AU in a disk
with $0.01$ \msuns within $100$ AU and a surface density profile that
falls off as $r^{-1/2}$, or to about 13 AU in the Minimum Mass Solar Nebula.
The initial dust-to-gas ratio is $0.01$. Using the radius of the planet's 
orbit as our unit distance, the inner boundary is at $r=0.4$ 
and the outer boundary at $r=2.5$. The gas rotates with a slightly 
sub-Keplerian velocity due to the radial pressure gradient. The dust 
rotates exactly with the Keplerian velocity, initially. All radial 
velocities are taken to be zero. The constant kinematic viscosity is set so 
that the \cite{1973A&A....24..337S} $\alpha$ parameter equals $0.004$ at $r=1$.
For a disk described above, the critical planet mass for gap opening is about 
$1$ Jupiter mass ($1$ \mjup) \citep{1999ApJ...514..344B}. We varied the mass 
of the planet between 0.001 \mjups and 0.5 \mjup. The boundary conditions are 
as in \cite{paardekooper03}, deriving from \cite{1996MNRAS.282.1107G}. They 
are non-reflecting, so that all waves generated in the simulated region leave
the computational domain without further influencing the interior. 

\section{Results}
\label{secres}
The evolution of the gas component is described in detail in 
\cite{paardekooper03}, and the drag forces on the gas are too small to 
alter the flow. Therefore we focus on the evolution of the dust particles.

\subsection{General flow structure}
For the standard case we use a planet of 0.1 \mjup, and we
consider dust particles of 1 mm. In the left panel of Fig. \ref{fig1}
we can see that the dust particles are opening a gap, which leads to 
variations in the dust-to-gas ratio of more than an order of magnitude 
(middle panel). For this small planet these variations in dust density 
are so large that the spiral waves are no longer visible. 
The radial cut (right panel) shows that this planet is 
not able to open up a gap in the gas, while the dust reacts more strongly 
and shows a very dense ring at $r=1.3$ just outside a very low density gap.

Also shown in the right panel of Fig. \ref{fig1} are positions of the 
3:2, the 2:1 and the 3:1 mean motion resonances. Especially the 3:2
resonance (at $r=1.25$) plays a role in structuring the disk, while 
only faint features can be seen at the other resonances. There also
exists a small density bump at corotation ($r=1.0$). 

\begin{figure}
\resizebox{\hsize}{!}{\includegraphics[bb=30 5 285 300]{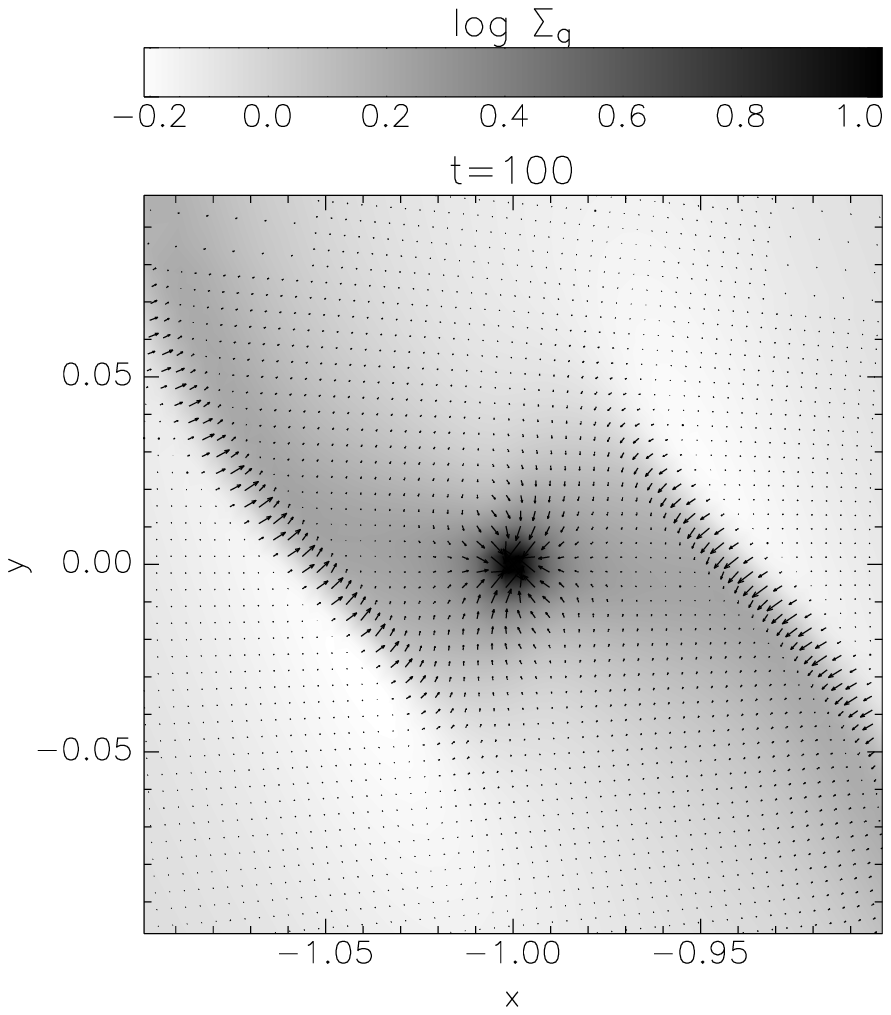}}
\caption{Close-up on the gas density near the planet for a high resolution
AMR simulation (3 levels of refinement, so a resolution that is 8 times higher 
than our standard resolution). The arrows indicate the relative velocity of 
the dust compared to the gas. The typical relative velocity across the shocks 
is 0.035 $c_{\mathrm{s}}$. Note that the color scale is reversed with respect 
to Fig. \ref{fig1}.}
\label{fig2}
\end{figure}

Figure \ref{fig2} shows a close-up on the gas density near the planet
in a high-resolution simulation. The high-density core can be seen, as
well as the spiral shocks. The arrows indicate the relative velocity
of the dust compared to the gas, showing that the dust decouples from
the gas near the spiral shocks. In the shocks the gas velocity is
reduced, and the gas is also deflected. The dust particles do not feel
the shock directly, but notice its presence by the drag forces
in the post-shock region. Figure \ref{fig2} shows that it takes a
certain distance before the velocity difference between dust and gas
has disappeared again. Because of this, the flux of dust particles
streaming into the spiral wave is larger than in the case of a
perfectly coupled gas-dust mixture. Since the wave is responsible for
gradually pushing the gas and dust particles away from the orbit of
the planet this leads to a depletion of dust particles, and hence a
deeper gap in the dust distribution. The dust density is affected most
where the waves are the strongest, which is approximately at a radial
distance $h$ from the orbit of the planet. This is where a low density
gap starts to form, leaving the particles at $r=1$ largely
unaffected. The particles moving to larger radii seem to get trapped in the
3:2 and the 2:1 mean motion resonances, creating the dense rings at
$r=1.3$ and $r=1.6$. Interestingly, it has been found that the Kuiper Belt 
Objects also tend to accumulate at the the 3:2 and 2:1 mean motion resonances
with Neptune (Luu \& Jewitt 2002). Whether this is related to the
effect we find, requires further study. Inside corotation, the combined 
inward force due
to the spiral waves, the imposed radial temperature gradient and the
viscous gas flow is so large that basically all particles move over
the resonances quickly and end up near the wave-damping boundary where they
are absorbed completely, leaving an almost featureless inner disk. 
Simulations with different planetary masses showed that the minimum planetary 
mass for this mechanism to operate is $0.05$ \mjups with particles of 1 mm. 

\subsection{Simulated observations}
\label{secobs}
\begin{figure*}
\includegraphics[bb=0 5 510 240,width=17cm]{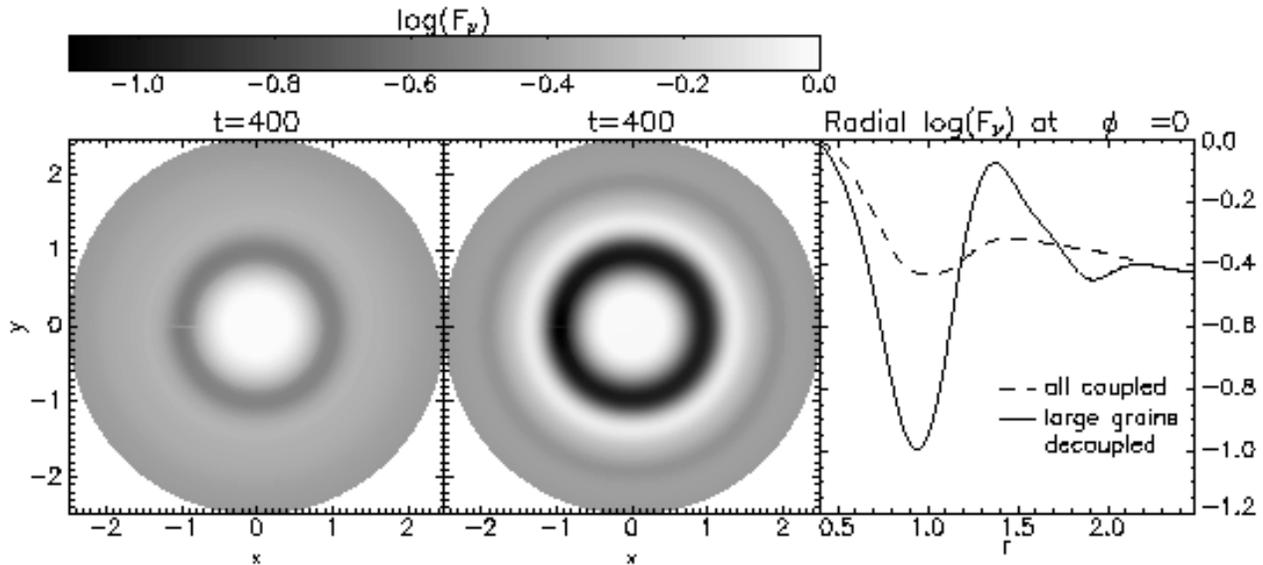}
\caption{Logarithm of flux densities at 1 mm, normalized by the maximum and 
convolved with a Gaussian of FWHM 2.5 AU, corresponding to a resolution of 
12 mas at 140 pc.
Left panel: all particles follow the gas exactly (static dust evolution). 
Middle panel: particles larger than the critical size decouple from the gas 
(dynamic dust evolution). Right panel: the corresponding radial flux 
densities.}
\label{fig3}
\end{figure*}

To investigate the observational effects of the gas-dust decoupling we 
simulate an image taken at a wavelength of 1 mm. The disk is optically thin at 
this wavelength, and we put the planet at $5.2$ AU. Calculations with different particle sizes showed that the minimum grain size
for gap formation is approximately $0.1$ mm, with a very sharp transition. 
We therefore assume that all particles larger than this critical size  create 
a gap, and smaller particles follow the gas exactly. To estimate the relative 
amount of mass in these two families of particles we take an MRN size 
distribution \citep{1977ApJ...217..425M}, in which the number density of 
particles is a power law in $s$ with exponent $-3.5$. For a maximum particle 
size of $5$ mm most mass is in particles larger than our critical size, and as 
a conservative assumption we take an equal amount of mass in the particles 
smaller than $0.1$ mm and in the larger particles. But since the opacity at 1 
mm is dominated by particles with $s\approx 1$ mm \citep{1993Icar..106...20M},
the emission will be dominated by the larger particles. We used the size-
dependent opacity data from \cite{1993Icar..106...20M} to calculate the flux
densities. In Fig. \ref{fig3} the resulting flux densities are shown, convolved with
a Gaussian of FWHM $2.5$ AU, which corresponds to an angular resolution of 
$12$ mas at 140 pc (the Taurus star forming region). This resolution is 
comparable to the maximum resolution of 10 mas that ALMA will achieve. In the 
left panel of Fig. \ref{fig3} we computed the flux under the assumption that 
all particles move with the gas, as was done for example by 
\cite{2002ApJ...566L..97W}. There is a little dip visible due to the decrease 
in density in the gas, but only when we include the density distribution of 
Fig. \ref{fig1} for the larger particles a clear gap emerges (middle panel 
of Fig. \ref{fig3}). The right panel of Fig. \ref{fig3} shows the radial 
dependence of the flux density. From this panel we can see that the contrast 
between the inside of the gap and the gap edges is enhanced from almost a 
factor of two (dashed line) to more than an order of magnitude (solid line). 
So the impact of the dynamics of the larger particles on the appearance of a 
low-mass planet in a disk is considerable.

\section{Summary}
\label{secdisc}
We have presented the first two-fluid calculations of a circumstellar disk 
with an embedded planet. These models allow us to make predictions on the dust 
distributions that will be observed in protoplanetary disks in the near 
future.

For a reasonable disk mass ($0.01$ \msuns within 100 AU) the strong spiral 
shocks near the planet are able to decouple the larger
particles ($\geq 0.1$ mm) from the gas. This leads to the formation of an
annular gap in the dust, even if there is no gap in the gas density. Because
the opacity at millimeter wavelengths is dominated by these larger particles,
the signatures of low-mass planets in disks can be stronger than previously 
thought. The minimum mass for a planet to open a gap this way was found to
be $0.05$ \mjups for 1 mm particles.

\begin{acknowledgements}
We wish to thank Carsten Dominik for carefully reading the manuscript. S.P. 
and G.M. acknowledge support from the European Research 
Training Network ``The Origin of Planetary Systems'' (PLANETS, contract number 
HPRN-CT-2002-00308).
\end{acknowledgements}

\bibliographystyle{aa} 
\bibliography{planet1.bib}

\end{document}